\documentclass[pre,amsmath,preprintnumbers,nofootinbib]{revtex4}
\usepackage{amsmath,amsfonts,bm}
\usepackage[dvips]{graphicx}

\begin{document}

\title{Comparison of far-from-equilibrium work relations}

\author{Christopher Jarzynski}

\affiliation{Department of Chemistry and Biochemistry, and Institute for Physical Science and Technology\\
University of Maryland, College Park, MD 20742 USA}

\date{\today}

\begin{abstract}
Recent theoretical predictions and experimental measurements have demonstrated
that equilibrium free energy differences can be obtained from
exponential averages of nonequilibrium work values.
These results are similar in structure, but not equivalent,
to predictions derived nearly three decades ago by Bochkov and Kuzovlev,
which are also formulated in terms of exponential averages but do not
involve free energy differences.
In the present paper the relationship between these two sets of results
is elucidated, then illustrated with an undergraduate-level solvable model.
The analysis also serves to clarify the physical interpretation of different
definitions of work that have been used in the context of thermodynamic
systems driven away from equilibrium.
\end{abstract}


\maketitle

\section{Introduction}

In recent years there has been considerable interest in the nonequilibrium
statistical mechanics of small systems~\cite{05BLR}.
Among the results that have been derived and tested experimentally,
the nonequilibrium work theorem~\cite{CJ97a,CJ97b},
\begin{equation}
\label{eq:nwt_intro}
\left\langle e^{-\beta W} \right\rangle = e^{-\beta\Delta F} ,
\end{equation}
relates fluctuations in the work $W$ performed during a thermodynamic
process in which a system is driven away from equilibrium,
to a free energy difference $\Delta F$ between two equilibrium states
of the system.
Here, $\beta$ specifies an inverse temperature,
and the angular brackets denote an average over an ensemble
of realizations (repetitions) of the process in question~\cite{pedagogical}.
Eq.~\ref{eq:nwt_intro} and closely related results~\cite{Cro99,Cro00,HuSz01}, along with
experimental confirmations~\cite{Liphardt02,Douarche05,Collin05,Blickle06},
have revealed that equilibrium free energy differences can be determined
from distributions of nonequilibrium work values.

The recent progress in this area has drawn attention to a set of earlier
papers by Bochkov and Kuzovlev~\cite{77BoKu,79BoKu,BK81a,BK81b},
in which the authors had obtained
-- as one consequence of a more general analysis --
the following result:
\begin{equation}
\label{eq:nwt0_intro}
\left\langle e^{-\beta W_0} \right\rangle = 1.
\end{equation}
The angular brackets and inverse temperature $\beta$ appearing here
have the same meaning as in Eq.~\ref{eq:nwt_intro},
and $W_0$ is identified as the work performed on the system.

Although Eqs.~\ref{eq:nwt_intro} and \ref{eq:nwt0_intro} are evidently similar
in structure, they are not identical;
most notably, $\Delta F$ does not appear,
either explicitly or implicitly\footnote{
Rewriting Eq.~\ref{eq:nwt_intro} in terms of {\it dissipated } work~\cite{CJ97a},
$W_d = W - \Delta F$, we obtain $\langle\exp(-\beta W_{\rm d})\rangle = 1$,
which bears an even stronger resemblance to Eq.~\ref{eq:nwt0_intro}.
However, the quantity $W_0$ appearing in Eq.~\ref{eq:nwt0_intro} is {\it not}
equivalent to $W_d$, as apparent from the definitions provided in Section~\ref{sec:setup}.
},
in Eq.~\ref{eq:nwt0_intro}.
The precise relationship between these two results
has not been clarified in the literature,
nor is it immediately obvious from a quick comparison of the original
derivations.
The aim of the present paper is to fill this gap, first by deriving the two
equalities within a single, Hamiltonian framework,
and then by illustrating them both using the simple model
of a perturbed harmonic oscillator.
The conclusions that will emerge from this analysis are summarized by
the following three points.
\begin{itemize}
\item
Eqs.~\ref{eq:nwt_intro} and \ref{eq:nwt0_intro} apply to the
same physical situation: a system, initially described
by an unperturbed Hamiltonian $H_0$, is driven away from
equilibrium by the application of a time-dependent perturbation.
In principle, a single set of experiments could be used to test both predictions.
\item
While both $W$ and $W_0$ are identified as {\it work} (in Refs.\cite{CJ97a,CJ97b}
and \cite{77BoKu,79BoKu,BK81a,BK81b}, respectively),
the two quantities generally differ;
see Eq.~\ref{eq:W} below.
The difference between them amounts to a matter of convention,
related to whether we choose to view
the perturbation as an external disturbance,
or else as a time-dependent contribution to the internal energy of the
system.
\item
For the special case of {\it cyclic} processes, in which the perturbation
is turned on and then off,
Eqs.~\ref{eq:nwt_intro} and \ref{eq:nwt0_intro} are equivalent.
\end{itemize}

This paper is organized as follows.
Section~\ref{sec:setup} establishes the Hamiltonian framework and
the notation that will be used throughout the paper.
In Section~\ref{sec:derivations} we derive Eqs.~\ref{eq:nwt_intro}
and \ref{eq:nwt0_intro} within this framework.
Section~\ref{sec:example} describes an exactly solvable model
-- a harmonic oscillator driven by a time-dependent external force --
that illustrates the validity of these predictions
and provides intuition regarding the two definitions
of work, $W$ and $W_0$.
Finally, Section~\ref{sec:weightedDists} presents an alternative derivation
of Eqs.~\ref{eq:nwt_intro} and \ref{eq:nwt0_intro}, by way of a stronger set of results
(Eq.~\ref{eq:weighted_summary}).
The paper concludes with a brief discussion.

\section{Setup}
\label{sec:setup}

To carry out a direct comparison between Eqs.~\ref{eq:nwt_intro} and \ref{eq:nwt0_intro},
we will use the setup considered in Ref.~\cite{BK81a}.
Consider a classical mechanical system with $D$ degrees of freedom,
described by coordinates ${\bf q} = (q_1,\cdots,q_D)$
and momenta ${\bf p} = (p_1,\cdots,p_D)$,
and let $z=({\bf q},{\bf p})$ denote a point in the phase space of this system.
Consider also a number of external forces $X_1, X_2, \cdots $,
which are under our direct control.
We act on the system by manipulating these forces.
The Hamiltonian that describes this system takes the form
\begin{equation}
\label{eq:defHam}
H(z;X) = H_0(z) - \sum_k X_k \, Q_k(z)
\end{equation}
(see Eq.~2.2 of Ref.~\cite{BK81a}), where
$Q_1(z), Q_2(z), \cdots$ denote the variables conjugate to the external forces:
\begin{equation}
\label{eq:QdHdX}
Q_k = - \frac{\partial H}{\partial X_k}.
\end{equation}
$H$ is a function on phase space, parametrized by
the forces $X =(X_1,X_2,\cdots)$.
We will refer to $H_0$ as the {\it bare}, or unperturbed, Hamiltonian,
and to $H$ as the {\it full} Hamiltonian.

If this system is brought into weak contact with a thermal reservoir at temperature $T$,
with the external forces held fixed,
then it will relax to an equilibrium state described by the Boltzmann-Gibbs distribution
\begin{equation}
\label{eq:bgdist}
P^{\rm eq}(z;X) = \frac{1}{Z(X)} \exp \left[ -\beta H(z; X) \right],
\end{equation}
where $\beta = (k_BT)^{-1}$.
The corresponding classical partition function and free energy are:
\begin{equation}
\label{eq:ZF}
Z(X) = \int dz \, \exp \left[ -\beta H(z; X) \right]
\qquad,\qquad
F(X) = -\beta^{-1} \ln Z(X) .
\end{equation}

Now imagine that we subject this system to a thermodynamic process,
defined by the following sequence of steps.
Prior to time $t=0$, the system is prepared in equilibrium,
in the absence of external forces, i.e.\ at
\begin{equation}
\label{eq:X0}
X_0 = (0,0,\cdots) .
\end{equation}
The reservoir is then removed.
Subsequently, from $t=0$ to a later time $t=\tau$, the external forces are
turned on according to some arbitrary but pre-determined schedule, or {\it protocol},
$X_t$.
The microscopic evolution of the system during this interval of time
is described by a trajectory $z_t$ evolving under
Hamilton's equations,
\begin{equation}
\label{eq:hameq}
\frac{d{\bf q}}{dt} = \frac{\partial H}{\partial{\bf p}}
\qquad,\qquad
\frac{d{\bf p}}{dt} = -\frac{\partial H}{\partial{\bf q}} ,
\end{equation}
where $H = H(z;X_t)$.
The protocol $X_t$ effectively traces out a curve in ``force space'',
from the origin (Eq.~\ref{eq:X0}) to some final point $X_\tau$.
Let $\Delta F$ denote the free energy difference between two equilibrium states
-- both at the same temperature $T$ --
associated with the initial and final forces:
\begin{equation}
\label{eq:DeltaF}
\Delta F = F(X_\tau) - F(X_0) =
-\beta^{-1} \ln \frac{Z(X_\tau)}{Z(X_0)}.
\end{equation}

By repeatedly subjecting the system to this process -- always first preparing
the system in equilibrium, and always using the same protocol $X_t$ --
we generate a number of statistically independent {\it realizations} of the process,
each characterized by a Hamiltonian trajectory $z_t$ describing the
microscopic response of the system to the externally imposed perturbation.
Angular brackets $\langle\cdots\rangle$ will specify
an ensemble average over such realizations.

For a given realization, let us now define
$W$ and $W_0$ appearing in Eqs.~\ref{eq:nwt_intro} and \ref{eq:nwt0_intro}:
\begin{subequations}
\begin{eqnarray}
\label{eq:W0def}
W_0 &=&  \int_0^\tau dt \,\sum_k X_k(t) \, \dot Q_k(z_t) \\
\label{eq:Wdef}
W &=& -  \int_0^\tau dt \, \sum_k \dot X_k(t) \, Q_k(z_t) ,
\end{eqnarray}
\end{subequations}
where the dots denote time derivatives,
e.g.\ $\dot Q_k = (d/dt) Q_k(z_t)$.
These two definitions are not equivalent: in general, $W \ne W_0$.

To gain some physical insight into these quantities, we rewrite them as follows:
\begin{subequations}
\label{eq:familiar}
\begin{eqnarray}
\label{eq:W0_familiar}
W_0 &=& \int dt \, X \cdot \dot Q =
\int X \cdot dQ  \\
\label{eq:W_familiar}
W &=& -\int dt \, \dot X \cdot Q = 
\int dX \cdot \frac{\partial H}{\partial X} ,
\end{eqnarray}
\end{subequations}
where $Q = (Q_1,Q_2,\cdots)$ is the vector of variables conjugate 
to the forces $X = (X_1,X_2,\cdots)$
(see Eq.~\ref{eq:QdHdX}).
The expression for $W_0$ is the familiar integral of force versus displacement
found in introductory textbooks on mechanics~\cite{HRW05},
and corresponds to the definition of work used by
Bochkov and Kuzovlev (Eq. 2.9 of Ref.~\cite{BK81a}).
By contrast, expressions equivalent to Eq.~\ref{eq:W_familiar} are often used to
define work in discussions of the microscopic foundations of macroscopic
thermodynamics~\cite{Gibbs,Schrodinger62,UhlenbeckFord};
this is the definition that is used in the context of
nonequilibrium work theorems (e.g.\ Eq.~3 of Ref.~\cite{CJ97a}).
While it might seem unusual that two different quantities, $W_0$ and $W$,
can both be interpreted as the work performed on a system,
this ambiguity simply reflects the freedom we have to define
what we mean by the internal energy of the system of interest.
We discuss this point in some detail in the following two paragraphs.

What is the internal energy of the system when its microstate is $z=({\bf q},{\bf p})$,
and the external forces are set at values $X = (X_1, X_2,\cdots)$?
Eq.~\ref{eq:defHam} suggests two natural ways to answer this question.
(i) We can take the internal energy to be given by the value of the bare Hamiltonian,
$H_0(z)$.
From this perspective the system is imagined
as a particle in a fixed energy landscape, $H_0$;
we affect the particle's energy by varying the forces so as to move it from
one region of phase space to another, but the forces $X$ do not themselves
appear in the definition of its energy.
(ii) Alternatively, we can define the internal energy to be given by the value
of the full Hamiltonian, $H = H_0 - X\cdot Q$.
This point of view is captured by imagining an energy landscape
that is not fixed, but changes with time as we manipulate the forces $X$.
Let us refer to these two alternatives as the (i) {\it exclusive} and the
(ii) {\it inclusive} frameworks, according to whether the term $-X\cdot Q$
is treated as a component of the internal energy of the system.

Now we use the Hamiltonian identity
\begin{equation}
\frac{\partial H}{\partial z} \cdot \frac{dz}{dt} =
\frac{\partial H}{\partial {\bf q}} \cdot \frac{\partial H}{\partial {\bf p}} -
\frac{\partial H}{\partial {\bf p}} \cdot \frac{\partial H}{\partial {\bf q}} = 0
\end{equation}
(see Eq.~\ref{eq:hameq}) to obtain
\begin{equation}
\frac{d}{dt} H(z_t;X_t) = 
\frac{\partial H}{\partial z} \cdot \frac{dz}{dt} +
\frac{\partial H}{\partial X} \cdot \frac{dX}{dt}
= \dot X \cdot \frac{\partial H}{\partial X} = -\dot X \cdot Q ,
\end{equation}
and therefore
\begin{equation}
\frac{d}{dt} H_0(z_t) = 
\frac{d}{dt}
\left(
H + X\cdot Q
\right) = X\cdot\dot Q.
\end{equation}
Comparing with Eq.~\ref{eq:familiar}, we see that $W_0$ and $W$ are equal to
the net changes in the values of $H_0$ and $H$, respectively,
during the interval of perturbation:
\begin{subequations}
\label{eq:W}
\begin{eqnarray}
\label{eq:W0dH0}
W_0 &=& \int_0^\tau dt \, \frac{dH_0}{dt} = H_0(z_\tau) - H_0(z_0) \\
\label{eq:WdH}
W &=& \int_0^\tau dt \, \frac{dH}{dt} = H(z_\tau;X_\tau) - H(z_0;X_0).
\end{eqnarray}
\end{subequations}
Since the system is thermally isolated (i.e.\ not in contact with a heat
reservoir) from $t=0$ to $t=\tau$,
it is natural to identify the work performed on it with the net change in its internal energy.
With this in mind, Eq.~\ref{eq:W} provides a simple interpretation of
the difference between $W$ and $W_0$.
If we adopt the exclusive point of view and take
the internal energy to be the value of the bare Hamiltonian $H_0$,
then $W_0$ is the work performed on the system, by the application of external
forces that affect its motion in a fixed energy landscape.
If we instead choose the inclusive framework, using
the full Hamiltonian $H = H_0 - X\cdot Q$
to define the internal energy of the system,
then $W$ is the appropriate definition of work.
The distinction between these two frameworks is illustrated with a specific
example in Section~\ref{sec:example}.

From Eq.~\ref{eq:W} we obtain an explicit expression for the difference between
$W$ and $W_0$:
\begin{equation}
\label{eq:Wdiff}
W_0 - W
= X_\tau \cdot Q(z_\tau)
= \sum_k X_k(\tau) \, Q_k(z_\tau),
\end{equation}
since $X_0=(0,0,\cdots)$.

\section{Derivations}
\label{sec:derivations}

Let us now compute the averages
of $e^{-\beta W_0}$ and $e^{-\beta W}$, over an ensemble of realizations
of the thermodynamic process described above.
Since the system evolves under deterministic (Hamiltonian) equations
of motion from $t=0$ to $t=\tau$, a given realization
is uniquely determined by the initial conditions $z_0$.
We can therefore express
$\langle e^{-\beta W_0}\rangle$ as an integral over an equilibrium distribution
of initial conditions:
\begin{equation}
\label{eq:1st.step}
\left\langle e^{-\beta W_0} \right\rangle =
\int dz_0\, 
P^{\rm eq}(z_0;X_0) \, e^{-\beta W_0(z_0)},
\end{equation}
where $W_0(z_0)$ denotes the value of $W_0$
for the trajectory launched from the microstate $z_0$.
The first factor in the integrand is
\begin{equation}
P^{\rm eq}(z_0;X_0) = \frac{1}{Z(X_0)} \,
e^{-\beta H_0(z_0)}
\end{equation}
[note that $H(z_0;X_0) = H_0(z_0)$, by Eq.~\ref{eq:X0}].
Using Eq.~\ref{eq:W0dH0}, we have
\begin{equation}
\label{eq:W0.explicit}
W_0(z_0) = H_0\left( z_\tau(z_0) \right) - H_0(z_0),
\end{equation}
where $z_\tau(z_0)$ indicates the final microstate of this trajectory,
expressed as an explicit function of the initial microstate.
Upon substituting these expressions into Eq.~\ref{eq:1st.step},
a cancellation of terms occurs in the exponents, and we get
\begin{equation}
\left\langle e^{-\beta W_0} \right\rangle =
\frac{1}{Z(X_0)}
\int dz_0\, e^{-\beta H_0(z_\tau(z_0))}.
\end{equation}
Since there is a one-to-one correspondence between the initial and
final conditions of a given trajectory, we can change the variables
of integration from $z_0$ to $z_\tau$:
\begin{equation}
\left\langle e^{-\beta W_0} \right\rangle =
\frac{1}{Z(X_0)}
\int dz_\tau
\left\vert \frac{\partial z_\tau}{\partial z_0} \right\vert^{-1}
\, e^{-\beta H_0(z_\tau)} .
\end{equation}
We have inserted the determinant
of the Jacobian matrix associated with this change of variables.
By Liouville's theorem, this factor is identically unity,
$\vert \partial z_\tau / \partial z_0\vert = 1$,
which finally gives us
\begin{equation}
\label{eq:nwt0}
\left\langle e^{-\beta W_0} \right\rangle =
\frac{1}{Z(X_0)}
\int dz_\tau \, e^{-\beta H_0(z_\tau)}  = 1,
\end{equation}
by Eq.~\ref{eq:ZF}.

The exponential average of $W$ (rather than $W_0$) follows from
similar manipulations:
\begin{eqnarray}
\left\langle e^{-\beta W} \right\rangle &=&
\int dz_0\, 
P^{\rm eq}(z_0;X_0) \, e^{-\beta W(z_0)} \nonumber \\
&=& \frac{1}{Z(X_0)}
\int dz_0\, e^{-\beta H(z_\tau(z_0);X_\tau)} \nonumber \\
\label{eq:nwt}
&=& \frac{1}{Z(X_0)} \int dz_\tau\, e^{-\beta H(z_\tau;X_\tau)}
=
\frac{Z(X_\tau)}{Z(X_0)} =
e^{-\beta\Delta F} .
\end{eqnarray}
We have used
$W(z_0) = H(z_\tau(z_0);X_\tau) - H(z_0;X_0)$ (Eq.~\ref{eq:WdH})
to get from the first line to the second,
and a change of variables, $z_0 \rightarrow z_\tau$, to get to the third.

Eq.~\ref{eq:nwt0} was originally obtained by
Bochkov and Kuzovlev~\cite{77BoKu,79BoKu,BK81a,BK81b},
whereas Eq.~\ref{eq:nwt} is the nonequilibrium work theorem of Refs.~\cite{CJ97a,CJ97b}.
These results apply to two physically distinct quantities,
$W_0$ and $W$, corresponding to different conventions
for defining the internal energy of the system.
In each case the exponential average of work reduces to a ratio of partition functions.
In Eq.~\ref{eq:nwt0} the ratio is
$Z(X_0)/Z(X_0)$, i.e.\ unity;
while in Eq.~\ref{eq:nwt} it is
$Z(X_\tau)/Z(X_0)$, which yields the free energy difference $\Delta F$.

Let us now consider the special case in which the external forces vanish
both at $t=0$ and at $t=\tau$:
\begin{equation}
X_0 = X_\tau = (0,0,\cdots).
\end{equation}
This corresponds to a {\it cyclic} process, for
which the Hamiltonian begins and ends at $H_0$.
In this case we have, identically,
$W=W_0$ (Eq.~\ref{eq:Wdiff}) and $\Delta F=0$ (Eq.~\ref{eq:ZF}).
Thus, Eqs.~\ref{eq:nwt0} and \ref{eq:nwt} are equivalent
when the Hamiltonian is varied cyclically.

Finally, it is instructive to consider a process during which
the external forces are switched on {\it suddenly} at $t=0$,
from $X_0=(0,0,\cdots)$ to  $X_\tau=(X_1,X_2,\cdots)$.
Since the process occurs instantaneously ($\tau\rightarrow 0$),
the system has no opportunity to evolve, hence $z_\tau=z_0$.
Thus, Eq.~\ref{eq:W} gives us
\begin{equation}
W_0 = 0
\qquad,\qquad
W = \Delta H(z_0),
\end{equation}
where $\Delta H(z) \equiv H(z;X_\tau) - H(z;X_0)$.
Eq.~\ref{eq:nwt0} is immediately satisfied, and
Eq.~\ref{eq:nwt} reduces to Zwanzig's perturbation formula~\cite{zwanzig},
\begin{equation}
\langle e^{-\beta\Delta H}\rangle_0 = e^{-\beta\Delta F},
\end{equation}
where $\langle\cdots\rangle_0$ denotes an average over microstates sampled
from the $X = (0,0,\cdots)$ canonical distribution.

\section{Example}
\label{sec:example}

\begin{figure}[htbp] 
   \centering
   \includegraphics[width=4.5in,angle=-90]{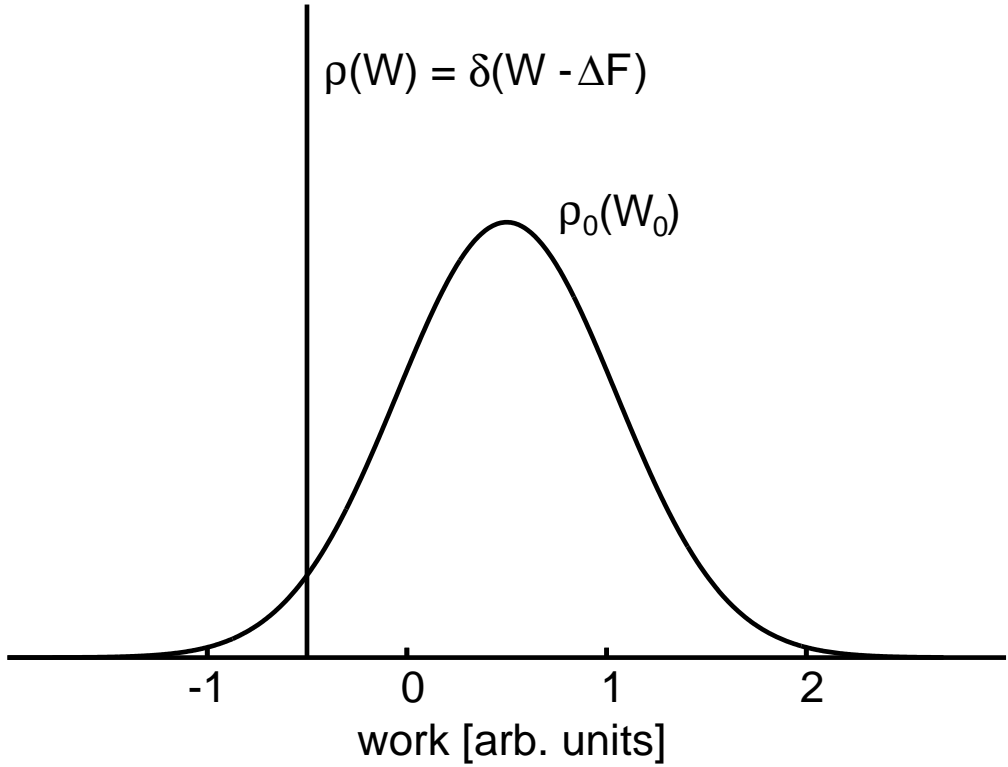} 
   \caption{Distributions of work values $W_0$ and $W$ for the harmonic oscillator
   example, with $\chi=m=\omega = 1.0$ and $k_BT = 0.3$.
   The distribution
   $\rho$ is a delta-function at $\Delta F = -0.5$, while
   $\rho_0$ is a Gaussian whose mean is at $-\Delta F$.}
   \label{fig:distsW}
\end{figure}

Let us now illustrate the general analysis presented above,
using the example of a one-dimensional harmonic oscillator
perturbed by a uniform external force.
We take the bare Hamiltonian
\begin{equation}
H_0(z) = \frac{1}{2m} p^2 + \frac{m\omega^2}{2} q^2,
\end{equation}
and we consider a perturbation
\begin{equation}
-X Q(z) = -X q.
\end{equation}
Thus, $H = H_0 - Xq$.
The perturbation describes a force $X$ acting along the
direction of the coordinate $q$.
The canonical distribution at a given force  $X$ is
\begin{equation}
P^{\rm eq}(z;X) = \frac{1}{Z(X)}
\exp\left[-\beta(H_0-X q)\right],
\end{equation}
and by direct evaluation of Eq.~\ref{eq:ZF} we get 
\begin{equation}
F(X) = F(0) -\frac{X^2}{2m\omega^2}.
\end{equation}
Now imagine a process during which the perturbing force is linearly ramped up
from zero to some positive value $\chi$:
\begin{equation}
X_t = \frac{\chi t}{\tau} 
\quad,\quad
0 \le t \le\tau.
\end{equation}
To simplify the calculations below, we take $\tau$ to be
the period of the unperturbed oscillator:
\begin{equation}
\label{eq:tau}
\tau = \frac{2\pi}{\omega}.
\end{equation}
The evolution of the system satisfies Hamilton's equations,
\begin{equation}
\dot q = \frac{p}{m}
\qquad,\qquad
\dot p = -m\omega^2 q + \frac{\chi t}{\tau} ,
\end{equation}
which can readily be solved.
For initial conditions $(q_0,p_0)$, we get a trajectory
\begin{subequations}
\label{eq:trajectory}
\begin{eqnarray}
q_t &=& q_0 \cos(\omega t) + \frac{p_0}{m\omega} \sin(\omega t)
+ \frac{\chi }{m\omega^3\tau} \left[ \omega t - \sin (\omega t) \right ] \\
p_t &=& p_0 \cos(\omega t) - m\omega q_0 \sin(\omega t)
+ \frac{\chi }{\omega^2\tau} \left[ 1 - \cos(\omega t) \right],
\end{eqnarray}
\end{subequations}
hence
\begin{equation}
\label{eq:finalmicrostate}
q_\tau = q_0 + \frac{\chi }{m\omega^2}
\qquad,\qquad
p_\tau = p_0.
\end{equation}
The quantities $W_0$ and $W$ then follow from Eq.~\ref{eq:W}:
\begin{equation}
\label{eq:W0W.ho}
W_0 = \chi q_0 - \Delta F
\qquad,\qquad
W = \Delta F,
\end{equation}
where
\begin{equation}
\Delta F = F(\chi)-F(0) = -\frac{\chi^2}{2m\omega^2}.
\end{equation}

From Eq.~\ref{eq:W0W.ho} we obtain explicit expressions for the
distributions of work values, $\rho_0(W_0)$ and $\rho(W)$,
assuming initial conditions $(q_0,p_0)$ sampled from equilibrium.
Since $W = \Delta F$ for every realization, we have
\begin{equation}
\rho(W) = \delta \left( W -\Delta F \right) .
\end{equation}
In turn, since $W_0$ is a linear function of $q_0$ (Eq.~\ref{eq:W0W.ho}),
which is sampled from a thermal, Gaussian distribution
with mean $\langle q_0\rangle=0$
and variance $\sigma_{q_0}^2 = (m\omega^2\beta)^{-1}$,
it follows that $W_0$ is also distributed as a Gaussian,
with mean and variance
$\langle W_0\rangle = -\Delta F$, 
$\sigma_{W_0}^2 = \chi^2 \sigma_{q_0}^2$.
Explicitly,
\begin{equation}
\rho_0(W_0) =
\sqrt{\frac{m\omega^2\beta}{2\pi\chi^2}}
\exp \left[
-\frac{m\omega^2\beta}{2\chi^2}
\left( W_0 + \Delta F \right) ^2 \right].
\end{equation}
It is now straightforward to verify by inspection and Gaussian integration
that Eqs.~\ref{eq:nwt_intro} and \ref{eq:nwt0_intro} are satisfied:
\begin{eqnarray}
\left\langle e^{-\beta W} \right\rangle &=&
\int dW \, \rho(W) \, e^{-\beta W} = e^{-\beta\Delta F} \\
\left\langle e^{-\beta W_0} \right\rangle &=&
\int dW_0 \, \rho_0(W_0) \, e^{-\beta W_0} = 1.
\end{eqnarray}

The very simple expressions obtained above for $W_0$ and $W$ are consequences of our
choice for $\tau$, Eq.~\ref{eq:tau}.
The model remains solvable for arbitrary $\tau$;
in that case both $\rho$ and $\rho_0$ are Gaussian distributions,
satisfying Eqs.~\ref{eq:nwt_intro} and \ref{eq:nwt0_intro},
but the expressions for their means and variances are more complicated.

\begin{figure}[htbp] 
   \centering
   \includegraphics[width=4.5in,angle=-90]{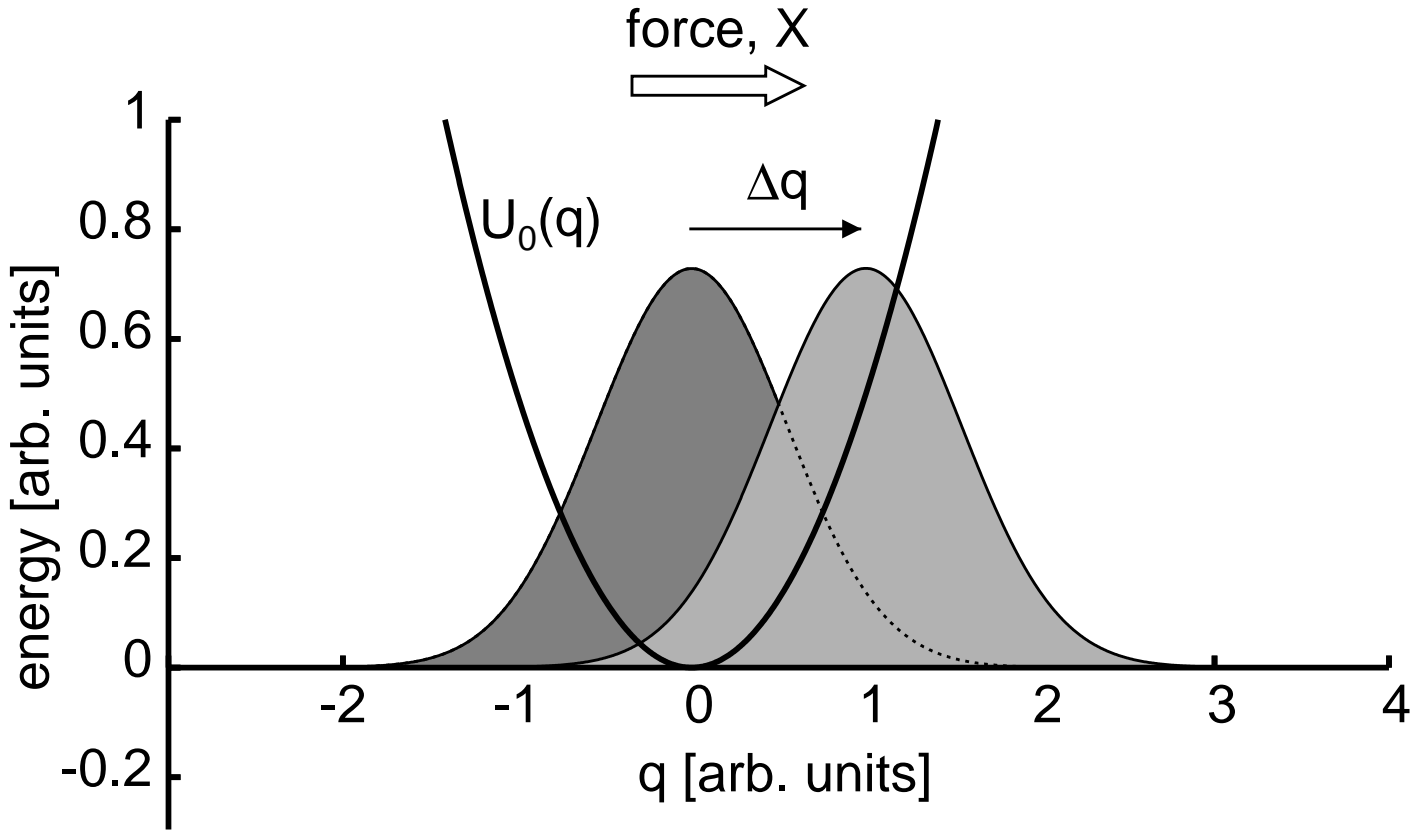} 
   \caption{The bare harmonic potential $U_0$ is shown, along with the distributions
   of initial particle positions (dark gray) and final particle positions (light gray).
   As we turn on the force $X$, we shift the distribution by an amount
   $\Delta q = \chi/m\omega^2 = 1.0$ away from the minimum of the potential,
   resulting in an average increase in the value of $U_0$.}
   \label{fig:excl}
\end{figure}

Fig.~\ref{fig:distsW} depicts the distributions $\rho_0$ and $\rho$.
Note that $W$ is negative, while the mean value of $W_0$ is positive.
We can understand this sign difference as follows.
Suppose we adopt the exclusive convention and
take the internal energy of the system to be given by $H_0$.
We imagine that the particle evolves in a fixed harmonic well
\begin{equation}
U_0(q) = m\omega^2 q^2/2,
\end{equation}
under the influence of a time-dependent external force $X_t$.
The initial position $q_0$ is sampled from a Gaussian distribution
centered at $q=0$,
and as we turn on the perturbing force from $0$ to $\chi$, we displace the
particle rightward by a net amount $\Delta q = q_\tau - q_0 = \chi/m\omega^2$
(Eq.~\ref{eq:finalmicrostate}).
The final condition $q_\tau$ is then distributed as a Gaussian
whose mean no longer coincides with the center of the fixed harmonic well,
but rather has shifted by a distance $\Delta q$, as shown in Fig.~\ref{fig:excl}.
In effect, the perturbation pushes the particle distribution rightward along the $q$-axis,
and ``up'' the quadratic potential energy landscape, resulting in a positive value for
the average work, $\langle W_0\rangle > 0$.

\begin{figure}[htbp] 
   \centering
   \includegraphics[width=4.5in,angle=-90]{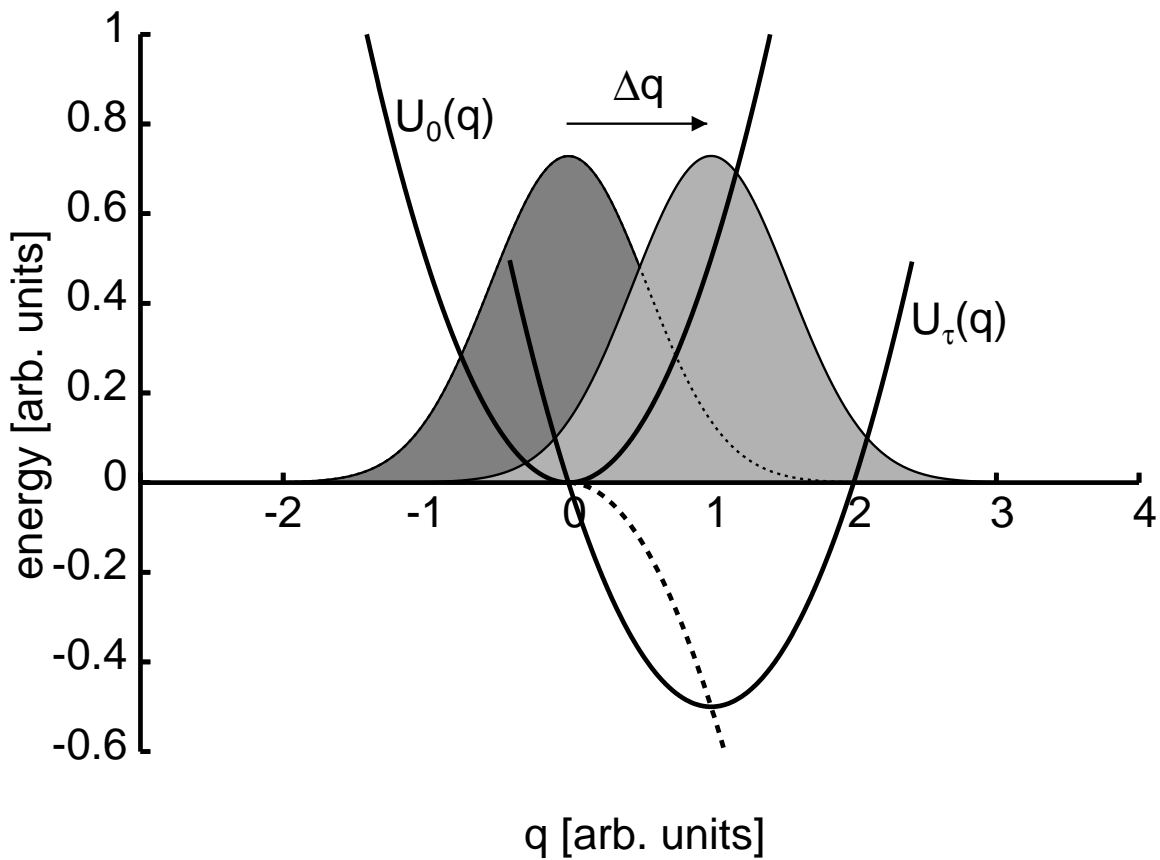} 
   \caption{Same as Fig.~\ref{fig:excl}, only now we think in terms of a time-dependent
   potential well rather than an externally applied force.
   From $t=0$ to $t=\tau$ the minimum of the harmonic well moves
   along the path depicted by the thick dashed line.
   During a given realization, the particle suffers a net displacement
   $q_\tau - q_0 = \Delta q$, hence $W = U_\tau(q_\tau) - U_0(q_0) = \Delta F$.}
   \label{fig:incl}
\end{figure}

Now suppose that we instead choose the inclusive convention and
use the full Hamiltonian $H = H_0 - Xq$ to define the internal energy of the system.
Thus we imagine a particle moving in a time-dependent potential,
\begin{equation}
\label{eq:Ut}
U_t(q) = U_0(q) - X_t q =
\frac{m\omega^2}{2} 
\left( q - \frac{\Delta q}{\tau} \, t \right)^2
+ \Delta F \cdot \frac{t^2}{\tau^2} ,
\end{equation}
where $\Delta q = \chi/m\omega^2$ and $\Delta F = - \chi^2/2m\omega^2$,
as above.
Eq.~\ref{eq:Ut} describes a harmonic well that moves rightward along the
$q$-axis with a velocity $\Delta q/\tau$, and slides downward in energy,
as depicted in Fig.~\ref{fig:incl}.
We can now appreciate why $W = \Delta F$ for every realization
of the process.
From $t=0$ to $t=\tau$ the particle moves by a net amount $\Delta q$;
simultaneously, the well shifts by the same amount,
and acquires an energy offset $\Delta F$:
\begin{equation}
U_\tau(q) = \frac{m\omega^2}{2} (q-\Delta q)^2 + \Delta F .
\end{equation}
The particle thus ends with the same displacement relative to the
minimum of the well as it began with,
so the net change in its energy is just the offset $\Delta F$.

In summary, in the exclusive framework,
we picture a particle that is pushed rightward by an external force
in a fixed harmonic well
($\langle W_0\rangle > 0$);
while in the inclusive framework
we imagine a particle that is dragged rightward in space and
``downward'' in energy by a moving harmonic well
($W < 0$).

\section{Weighted Distributions}
\label{sec:weightedDists}

Here we sketch
an alternative derivation of Eqs.~\ref{eq:nwt_intro} and \ref{eq:nwt0_intro}.

Consider an ensemble of realizations of the process described in Section~\ref{sec:setup}.
Let us picture this ensemble as a swarm of Hamiltonian trajectories evolving in phase space,
represented by a density~\footnote{
See Ref.~\cite{Ladek} for a brief discussion of the ordering of limits implied in Eqs.~\ref{eq:fdef}
and \ref{eq:g0def}.}
\begin{equation}
\label{eq:fdef}
f(z,t) = \Bigl\langle \delta\left( z - z_t \right) \Bigr\rangle ,
\end{equation}
which satisfies the Liouville equation,
\begin{equation}
\label{eq:liouville}
\frac{\partial f}{\partial t}
= \{H,f\} .
\end{equation}
Here we use the convenient Poisson bracket notation:
$\{A,B\} = ({\partial A}/{\partial{\bf q}}) \cdot ({\partial B}/{\partial{\bf p}}) -
({\partial A}/{\partial{\bf p}}) \cdot ({\partial B}/{\partial{\bf q}})$ .
In general, Eq.~\ref{eq:liouville} does not yield a simple solution;
the evolution of $f(z,t)$ can be very complicated, particularly if the underlying
Hamiltonian dynamics are chaotic.

For a given trajectory $z_t$, let
\begin{equation}
\label{eq:w0def}
w_0(t) = \int_0^t dt^\prime \, \sum_k X_k(t^\prime)\,\dot Q_k(z_{t^\prime})
\end{equation}
denote the amount of work performed on the system to time $t$,
using the definition of work corresponding to the exclusive
framework (Eq.~\ref{eq:W0def} and Refs.~\cite{77BoKu,79BoKu,BK81a,BK81b}).
Since the rate of change of the observable
$Q_k$ along a trajectory $z_t$ is given by $\dot Q_k = \{Q_k,H\}$~\cite{Goldstein.9.5},
we can rewrite Eq.~\ref{eq:w0def} as 
\begin{equation}
w_0(t) = \int_0^t dt^\prime \,  \{ X \cdot Q,H \} = \int_0^t dt^\prime \, \{ X \cdot Q,H_0 \} .
\end{equation}
The last equality follows from the identity $\{ X\cdot Q , X\cdot Q \} = 0$ .
Now consider a {\it weighted} phase space density
\begin{equation}
\label{eq:g0def}
g_0(z,t) = \Bigl\langle \delta\left( z - z_t \right) \, \exp[-\beta w_0(t)] \Bigr\rangle ,
\end{equation}
in which each trajectory carries a statistical weight,
$\exp[-\beta w_0(t)]$
(see the discussion below).
This density satisfies
\begin{equation}
\label{eq:g0_eom}
\frac{\partial g_0}{\partial t} = \{H,g_0\} - \beta  \{ X \cdot Q,H_0 \} g_0,
\end{equation}
where the second term on the right accounts for the evolving statistical weights.
The derivation of this equation is very similar to those found in Section~II of
Ref.~\cite{CJ97b} and Section~4.1 of Ref.~\cite{Ladek}.

Since $w_0(0) = 0$ identically,
and since we assume our ensemble is initially prepared in equilibrium,
we have
$g_0(z,0) = f(z,0) = P^{\rm eq}(z;X_0)$.
Given these initial conditions, the unique solution of Eq.~\ref{eq:g0_eom}
is the time-{\it independent} distribution
\begin{equation}
\label{eq:g0_soln}
g_0(z,t) = \frac{1}{Z(X_0)} \exp[-\beta H_0(z)] = P^{\rm eq}(z;X_0).
\end{equation}
To see this, note that
\begin{equation}
\{ H , e^{-\beta H_0} \} = -\beta \{ H , H_0 \} \, e^{-\beta H_0}
= \beta \{ X\cdot Q , H_0 \} e^{-\beta H_0} ,
\end{equation}
using the derivative rule for Poisson brackets,
and the identity $\{ H_0 , H_0 \} = 0$.
From this result it follows by inspection
that Eq.~\ref{eq:g0_soln} satisfies Eq.~\ref{eq:g0_eom}.

The functions $f(z,t)$ and $g_0(z,t)$ are two different statistical representations
of the same ensemble of realizations.
Continuing to picture this ensemble as a swarm of trajectories evolving
in phase space,
$f$ (Eq.~\ref{eq:fdef}) can be viewed as a number density,
which simply counts how many trajectories are found in the vicinity of $z$ at time $t$;
while $g_0$ (Eq.~\ref{eq:g0def}) can be interpreted as a mass density,
if we imagine that each realization carries a fictitious, time-dependent mass $\exp[-\beta w_0(t)]$.
Eq.~\ref{eq:g0_soln} then has the following interpretation:
when the initial conditions are sampled from equilibrium,
the ``mass density'' of trajectories remains constant in time, even as the ``number density" evolves
in a possibly complicated way.
Thus while the number of trajectories found
near a given point $z$ changes with time, these fluctuations are
balanced by the evolving statistical weights (fictitious masses) of those trajectories,
in precisely such a way as to keep the local mass density constant.

We can obtain analogous results 
in the inclusive framework (Eq.~\ref{eq:Wdef} and Refs.~\cite{CJ97a,CJ97b}).
Introducing
\begin{equation}
w(t) = - \int_0^t dt^\prime \, \sum_k \dot X_k(t^\prime)\, Q_k(z_{t^\prime})
= -\int _0^t dt^\prime \, \dot X \cdot Q,
\end{equation}
along with the corresponding weighted density
\begin{equation}
g(z,t) = \Bigl\langle \delta\left( z - z_t \right) \, \exp[-\beta w(t)] \Bigr\rangle ,
\end{equation}
we obtain the equation of motion~\cite{CJ97b,Ladek}
\begin{equation}
\label{eq:g_eom}
\frac{\partial g}{\partial t} = \{H,g\} + \beta  \dot X Q g .
\end{equation}
For initial conditions $g(z,0) = f(z,0) = P^{\rm eq}(z;X_0)$,
the unique solution is
\begin{equation}
\label{eq:g_soln}
g(z,t) = \frac{1}{Z(X_0)} \exp [-\beta H(z;X_t)] =
\frac{Z(X_t)}{Z(X_0)} \, P^{\rm eq}(z;X_t) .
\end{equation}
The weighted density is no longer constant in time (as was the case with $g_0$),
but rather is proportional to the equilibrium distribution corresponding to the current
value of the parameters $X$.

The results just obtained are summarized as follows:
\begin{subequations}
\label{eq:weighted_summary}
\begin{eqnarray}
\label{eq:weighted_g0}
\Bigl\langle \delta\left( z - z_t \right) \, \exp[-\beta w_0(t)] \Bigr\rangle 
&=& P^{\rm eq}(z;X_0) \\
\label{eq:weighted_g}
\Bigl\langle \delta\left( z - z_t \right) \, \exp[-\beta w(t)] \Bigr\rangle
&=& \frac{Z(X_t)}{Z(X_0)} \, P^{\rm eq}(z;X_t) .
\end{eqnarray}
\end{subequations}
Eqs.~\ref{eq:nwt_intro} and \ref{eq:nwt0_intro} now follow immediately
by evaluating Eq.~\ref{eq:weighted_summary} at $t=\tau$
and integrating both sides over phase space.
While the derivations presented here are less elementary than
those of Section~\ref{sec:derivations},
we ultimately gain a stronger set of results.
By a simple trick of statistical reweighting,
we transform an equation of motion that we cannot solve
(Eq.~\ref{eq:liouville})
into one that is easily solved (Eq.~\ref{eq:g0_eom} or \ref{eq:g_eom}).
The result, Eq.~\ref{eq:weighted_summary}, allows us to
reconstruct equilibrium distributions $P^{\rm eq}$
using trajectories driven away from equilibrium.

Eqs.~\ref{eq:weighted_g0} and \ref{eq:weighted_g} are in fact equivalent.
Multplying both sides of Eq.~\ref{eq:weighted_g0} by
$\exp[+\beta X_t \cdot Q(z)]$ and pulling this factor inside the angular brackets,
we obtain Eq~\ref{eq:weighted_g}.
Conversely, multiplication by $\exp[-\beta X_t \cdot Q(z)]$
leads us from Eq.~\ref{eq:weighted_g} to Eq.~\ref{eq:weighted_g0}.
However, this equivalence is lost once we integrate over phase space:
Eqs.~\ref{eq:nwt_intro} and \ref{eq:nwt0_intro} do not imply one another.

Eq.~\ref{eq:weighted_g} can be viewed as a direct consequence
of the Feynman-Kac theorem; this observation by Hummer and Szabo serves as
a starting point for their method of reconstructing equilibrium potentials of mean force from
single-molecule manipulation experiments carried out away from equilibrium~\cite{HuSz01}.
Moreover, Eq.~\ref{eq:weighted_g0} is essentially a special case of Eq.~12 of
Ref.~\cite{HuSz01} (with $W_0$ as generalized by Eq.~\ref{eq:work_nonlin} below),
if we assume that their confining potential is initially
turned off: $u(z,0) = 0$.
For an alternative approach to estimating potentials of mean force from
similar experiments, see the ``clamp-and-release'' method proposed by Adib~\cite{Adib06}.

\section{Discussion}

The nonequilibrium work theorem, Eq.~\ref{eq:nwt_intro}, has generated interest
(and controversy~\cite{CohenMauzerall04,CohenMauzerall05,Sung.condmat0506214v2})
primarily for two reasons.
First, along with the {\it fluctuation theorem} for entropy
production~\cite{93ECM,94ES,95GC,98Kur,99LS,FT},
it is one of relatively few equalities in statistical physics that apply to
systems far from thermal equilibrium.
Note that the term ``fluctuation theorem''
has also been used to specify a relation between the response
of a system to external perturbations,
and a correlation function describing fluctuations of the
unperturbed system~\cite{Hanggi78,HanggiThomas82}.
Second, Eq.~\ref{eq:nwt_intro} predicts that equilibrium free energy differences can
be determined from irreversible processes,
counter to expectations that irreversible work values can only place
{\it bounds} on $\Delta F$~\cite{RMA01}.
Eq.~\ref{eq:nwt0_intro} shares the first feature
-- it remains valid far from equilibrium -- but not the second;
it does not seem to be the case that $\Delta F$ can be determined
solely from a distribution of values of $W_0$.

A crucial distinction in this paper has been the difference
between the quantities $W$ and $W_0$.
The recognition that, in the literature, various meanings are assigned to the term {\it work},
might at first come as an unwelcome surprise.
Work is a concept of such central importance in thermodynamics
that it ought to be unambiguously defined!
However, in dealing with a physical situation that involves
the mechanical perturbation of a system,
the perturbation is inevitably accomplished by coupling
externally controlled variables ($X_1, X_2, \cdots$)
to generalized system coordinates ($Q_1, Q_2, \cdots$).
This coupling is represented by a term
of the form $-\sum_k X_k Q_k$ (or a nonlinear generalization thereof, see below)
in the full Hamiltonian that governs the evolution of the system and its surroundings.
We are then faced with the question of whether or not to view this term
as part of the internal energy of the system of interest.
As stressed in this paper, either choice is perfectly acceptable
-- this is a question of book-keeping rather than principle --
but it is precisely this freedom that leads to the ambiguity in the definition of work.
For related discussions of this issue, particularly in the context of interpretation
of experimental data, see Refs.~\cite{Sch03,Nar04,Dha05,Douarche05}.

Throughout this paper it has been assumed,
following Refs.~\cite{77BoKu,79BoKu,BK81a,BK81b},
that the coupling between the forces $X$ and the observables $Q$
is linear: $H = H_0 - X\cdot Q$.
However, as already observed by Bochkov and Kuzovlev, this assumption can
easily be relaxed.
Had we written the Hamiltonian as
\begin{equation}
H(z;X) = H_0(z) - h(Q;X) ,
\end{equation}
and assumed $h(Q;X_0)=0$,
then the entire analysis leading to Eqs.~\ref{eq:nwt_intro} and \ref{eq:nwt0_intro}
would have remained valid, provided the following definitions of work:
\begin{equation}
\label{eq:work_nonlin}
W = -\int_0^\tau dt \, \dot X \cdot \frac{\partial h}{\partial X}
\qquad,\qquad
W_0 = \int_0^\tau dt \, \dot Q \cdot \frac{\partial h}{\partial Q} .
\end{equation}
We recover Eqs.~\ref{eq:defHam} and \ref{eq:familiar} with a linear
perturbation $h = X\cdot Q$.

While the analysis here has been carried out using Hamiltonian dynamics,
the conclusions remain valid under other frameworks for modeling
the evolution of the system.
The connection to the stochastic approach taken in Ref.~\cite{HuSz01}
has already been noted.
Moreover, Eqs.~33 and 34 of Ref.~\cite{Nar04},
derived for inertial Langevin dynamics, are equivalent to Eqs.~\ref{eq:nwt_intro}
and \ref{eq:nwt0_intro} of the present paper.
For non-inertial (overdamped) Langevin dynamics, similar results follow
directly from the Onsager-Machlup expressions for path-space distributions~\cite{OM53,Dean}.

Finally, recall the Crooks fluctuation theorem~\cite{Cro99},
\begin{equation}
\label{eq:cft}
\frac{\rho_F(+W)}{\rho_R(-W)} = \exp[\beta(W-\Delta F)],
\end{equation}
where the subscripts refer to two thermodynamic process
({\it forward} and {\it reverse}) that are related by time-reversal of the protocol
used to perturb the system.
The Bochkov-Kuzovlev papers contain results
that are reminiscent of Eq.~\ref{eq:cft}, for instance Eq.~7 of Ref.~\cite{77BoKu}
and Eq.~2.12 of Ref.~\cite{BK81a}.
However, while Crooks uses a definition of work corresponding to $W$
of the present paper, Bochkov and Kuzovlev use $W_0$,
and their results do not involve $\Delta F$.
Moreover, in Refs.~\cite{77BoKu,79BoKu,BK81a,BK81b}
the derivations seem to assume that the initial conditions are sampled from
the same, unperturbed equilibrium distribution for both the forward and
the reverse process (see e.g.\ Eq.~2.6 of Ref.~\cite{BK81a}).
Crooks, by contrast, assumes that the forward and reverse
processes are characterized by different initial equilibrium states.
It would be useful to clarify more precisely the relationship between
these sets of results.

\section*{Acknowledgments}
It is a pleasure to acknowledge useful conversations and correspondence
with Artur Adib, R. Dean Astumian, Gavin Crooks, Abhishek Dhar, Peter H\" anggi,
Gerhard Hummer, and Attila Szabo;
and financial support provided by the University of Maryland (start-up research funds).

\end{document}